\documentstyle[preprint,aps,fleqn]{revtex} 

\newfont{\bi}{cmbxti10 scaled\magstep2}
\begin{document}
\draft

\title{Giant Low Temperature Heat Capacity of GaAs Quantum Wells near
Landau Level Filling $\nu$=1}
\author{V. Bayot, E. Grivei, S. Melinte}
\address{Unit\'e de Physico-Chimie et de Physique des
Mat\'eriaux,
Universit\'e Catholique de Louvain,\\
Place Croix du Sud 1, B-1348 Louvain-la-Neuve, Belgium}
\author{M.B. Santos and M. Shayegan}
\address{Department of Electrical Engineering, Princeton University,
Princeton N.J. 08544, USA}

\date{\today}
\maketitle

\begin{abstract}
We report low temperature ($T$) heat capacity ($C$) data on a 
multiple-quantum-well 
GaAs/AlGaAs sample in the quantum Hall regime.
Relative to its low field magnitude, $C$ exhibits up to
$\sim 10^5$-fold enhancement near $\nu$=1
where Skyrmions are
the ground state of the confined 
two-dimensional electrons.
We attribute the large $C$ to a Skyrmion-induced, strong coupling
of the nuclear spin system to the lattice.
The data are consistent with the Schottky nuclear heat capacity
of Ga and As atoms in the quantum
wells, except at very low $T$ where $C$ vs $T$ exhibits a
remarkably sharp peak suggestive of a {\it phase transition in the 
electronic system}.

\end{abstract}



Heat capacity is one of the most fundamental physical properties as
it directly probes thermodynamic quantities such as entropy \cite{Gopal66}.
In the case of two-dimensional electron systems (2DESs) heat capacity
can be a powerful probe of single- and many-body properties such
as the Landau quantized density of states and the quantum Hall effect
(QHE), both integral and fractional \cite{Prange87,HeatCap}.
Measurements of 2DES heat capacity, however, are among the
most challenging experiments because of the very small electron contribution.
As a result, in contrast to the overwhelming number of
magnetotransport and magnetooptical experiments reported for 2DESs
in the QHE regime \cite{2DESreview}, very few heat capacity
measurements have been reported so far \cite{Gornik85,Wang}.

In this Letter we report heat capacity measurements of a 
modulation-doped 
GaAs/Al$_{x}$Ga$_{1-x}$As multiple-quantum-well heterostructure
down to very low temperature ($T\geq$ 27 mK) and Landau
level filling factor ($\nu \gtrsim 0.5$).
As a function of increasing magnetic field, $B$, in addition
to oscillations associated with the 2DESs' oscillating density of states at
the Fermi level, we observe a dramatic increase of the low-$T$ heat
capacity ($C$)
in the range $0.5\lesssim \nu \lesssim 1.5$.
For $0.7 \lesssim \nu \lesssim 0.85$, $C$ exhibits a striking $T$-dependence, 
including a
remarkably sharp peak suggestive of a {\it phase transition} at very 
low $T$.
We interpret these unexpected observations in terms of the Schottky
model \cite{SchottkyRev} for the nuclear-spin heat capacity of Ga and As atoms
which couple to the lattice
via the 2DESs' low energy excitation spin textures (Skyrmions). The 
origin of the peak at very low $T$ is discussed in relation with a phase 
transition in the electronic system.


The experiments were performed on a multiple-quantum-well
heterostructure grown by molecular-beam epitaxy and consisting of 100 GaAs
quantum wells separated by Al$_{0.3}$Ga$_{0.7}$As barriers.
The wells and barriers are 250 and 1850 $\rm \AA$ thick respectively,
and the barriers are $\delta$-doped with donors (Si) near their centers.
Electrical resistivity data on samples from
the same wafer exhibit well-developed fractional QHE and attest to the high
quality of the sample \cite{bayot93}.
A 7 $\times$ 7 mm$^{2}$ piece of the wafer was
thinned to 65 $\mu$m and two carbon paint resistors were deposited on
the substrate side and connected to the heat sink with NbTi wires.
One carbon resistor was used as a thermometer while the other served
as a heater. The carbon thermometer was calibrated versus a 
$\rm{RuO_2}$ resistance
thermometer at $B=0$, and we checked that its calibration was negligibly 
affected by the magnetic field.
Depending on the external time constant of the sample $\tau_{ext}$, three
different techniques were used to measure $C$.
At low $B$, $\tau_{ext} \approx 0.1$ s and $C$ was measured
using the AC technique \cite{Sullivan68,Wang} at
26 Hz, a frequency near which $C$ was found to be frequency independent.
In the $B$ range near $\nu$=1, $C$ and hence $\tau_{ext}$
increased by up to five orders of magnitude and we turned to
a pulse technique \cite{pulseC,Gornik85}, either in the relaxation
regime ($C=\kappa \tau_{ext}$,
where $\kappa$ is the thermal conductance to the heat sink) or in the
quasi-adiabatic regime ($C=Q/\Delta T$ where $\Delta T$ is the temperature
increase resulting from the applied heat $Q$, after internal 
relaxation is completed), depending on $\tau_{ext}$. 
$C$ was measured in a dilution refrigerator while $B$ was applied 
either perpendicular to the 2DES plane ($\theta
=0^{\circ}$) or at an angle of $\theta =30\pm 2^{\circ}$. The 
absolute accuracy on the measured $C$ is of the order of $\pm$10
to 15$\%$, as illustrated in some of the figures.


The density of states at the Fermi level $D(E_{F})$ of a 2DES in a
perpendicular $B$ exhibits $1/B$-periodic oscillations due to the
formation of disorder-broadened Landau levels \cite{Prange87}.
$D(E_{F})$ is maximum at half-integral $\nu$ and is minimum at integral $\nu$.
The oscillating $D(E_{F})$ induces oscillations in many physical
properties such as electrical resistivity \cite{Prange87,2DESreview},
magnetization \cite{Eisenstein85} and electron specific heat:
$c_e=\pi^2/3 k_B^2TD\left({{E}_{F}}\right)$ (J/K) per electron  \cite{HeatCap}.
The data presented in Fig.~\ref{fig:1}(a) demonstrate the oscillatory behavior
of $C_e$ for a 2DES. Even though $C$ is dominated by the
lattice and addenda contributions, up to $\approx 10\%$ oscillations coming
from the 2DESs are clearly observed. Comparison with previous
data \cite{Wang} gives $D(E_{F})$ at
$B$=2.3T ($\nu$=5/2) about five times larger for our sample.
This is consistent with the lower disorder, and hence smaller width of Landau
levels, in our sample as evidenced by the
presence of minima in $C_e$ at odd $\nu$ down to $B$=1.14T ($\nu=5$). From 
the position of the minima, we infer a 2DES
density (per layer) of $n_{e}\approx 1.4\times 10^{11}$ cm$^{-2}$,
consistent with the magnetotransport data  \cite{bayot93}.

Near $\nu=1$, $C$ reveals a completely different and
unexpected behavior. Figure \ref{fig:1}(b) shows orders of magnitude
enhancement of $C$ with respect to the lower-$B$ data of Fig.~\ref{fig:1}(a).
$C$ exhibits large maxima at $\nu \approx
0.8$ and 1.2 and decreases rapidly for $\nu \gtrsim 1.2$ and $\nu
\lesssim 0.8$. Moreover, the $T$-dependence of $C$ is particularly striking 
(Fig.~\ref{fig:1c}): in contrast to the low-$B$ data
where $C$ {\it decreases} with decreasing $T$ (not shown here),
at $B$=7T
($\nu=0.81$) $C$ first {\it increases} with decreasing $T$
and then displays a very sharp peak at $T_{c} \approx$ 36 mK before
decreasing at very low $T$ \cite{peakC}. 
In the remainder of the paper, we concentrate on the unexpected high-$B$ data
near $\nu=1$.

We begin with a discussion of Fig.~\ref{fig:1c} data in the $T$
range $0.1 \lesssim T \lesssim 0.4$K where $C \propto T^{-2}$.
Both the very large magnitude of $C$ and the $T^{-2}$
dependence point to the nuclear
Schottky effect which results from the entropy reduction of the
nuclear spin system (NSS) with decreasing $T$ when the thermal energy $k_{B}T$
is much larger than the spin energy level spacing $\Delta$ \cite{SchottkyRev}.
The observation of the Schottky effect
requires good coupling of the NSS to the lattice in order to reach
thermal equilibrium in the time scale of the experiment.
This coupling is provided by electron
spin-flip excitations and further relaxation to the lattice \cite{Abragam78}.
Consequently, while the effect is commonly observed in metals, it
usually remains
undetected in high purity materials with low free-carrier density
\cite{Collan70}.
At first sight, because of their high purity and the low free-carrier
density in the quantum wells, GaAs/AlGaAs
heterostructures should not be good candidates for the observation of
a nuclear Schottky effect.
While this is supported by the low-$B$ data (Fig.~\ref{fig:1}(a))
where only the lattice, addenda and 2DESs contribute to the low value of $C$,
surprisingly a Schottky behavior is observed at higher $B$ near $\nu=1$. 
Why?

Recent theoretical \cite{Skyrme,Wu95,Brey95} and experimental
\cite{Tycko95,Barrett95,Schmeller95,Aifer96} work on 2DESs has shown that, 
in the limit of a
weak Zeeman coupling, electron spin textures known as {\it Skyrmions} are the
lowest energy, charged excitations of the ferromagnetic ground state 
near $\nu$=1 \cite{Prange87}. Skyrmions result from the dominance of the
Hartree energy over the Zeeman energy, prohibiting single spin-flip
excitations and favoring smooth distorsions of the spin field; the competition
between the two energies determines the total spin and size of Skyrmion
quasi-particles.

While the importance of the electron-nuclear spin interaction has been 
known for some time \cite{Barrett94,Vagner95}, recent nuclear magnetic 
resonance (NMR) 
experiments \cite{Tycko95,Barrett95} have provided clear evidence for a
strong coupling of the NSS to the lattice through finite-size
Skyrmions near $\nu=1$.
Subsequent theoretical results
\cite{Brey95} have shown excellent quantitative agreement with the NMR data
regarding
the spin polarization of the 2DES, $\langle S_{z} \rangle$, thereby
providing additional credance to the Skyrmionic picture around $\nu=1$.
Finally, very recently, further evidence for Skyrmions and estimation
of their size was reported from magnetotransport \cite{Schmeller95} 
and magnetooptical data \cite{Aifer96}.

Based on these observations, we
suggest that near $\nu$=1 Skyrmions induce a strong coupling of the NSS to the
lattice and the large nuclear heat capacity is observed.
The key role of Skyrmions is supported by: (1) the absence of the 
nuclear-spin contribution to our measured $C$ for $\nu \gtrsim 
1.5$ where the Skyrmions are no longer the ground state of the 2DES 
\cite{Wu95,Brey95,Schmeller95}, and (2) our experiments
in a tilted $B$, presented in Fig.~\ref{fig:2}, which
clearly show that the heat capacity anomaly relates to the 2DES filling
factor (rather than total $B$).

We now turn to a more quantitative interpretation of the data. Ga and
As have spin quantum number $I$=3/2. When $\Delta \ll k_{B}T$,
the (Schottky) nuclear specific heat is given by \cite{Gopal66,Girvin95}:
\begin{equation}
{c}_{N}={k}_{B}{5 \over 4}{\left({{\Delta  \over k_BT}}\right)}^{2}
\ \ \ \ \ \ \ \left({{J \over K\ \rm{nucleus}}}\right)
\label{CSchottky}
\end{equation}
where $\Delta = \alpha B$ and $\alpha / k_B = 4.9\times 10^{-4}, 6.23\times 
10^{-4}$ and $3.5\times 10^{-4}$ (K/T) for $^{69}$Ga (60.4$\%$), $^{71}$Ga
(39.6$\%$), and $^{75}$As (100$\%$), respectively \cite{Table}.
If we assume that only
the nuclear spins of the Ga and As atoms in the quantum wells
contribute to the observed $C$, we obtain for our sample a
nuclear heat capacity $C_N = 2.0\times 10^{-11} B^{2}\ T^{-2}$ (J/K) 
for $\Delta \ll k_{B}T$.
Figure \ref{fig:2}
clearly indicates that the calculated $C_{N}$ is semi-quantitatively
consistent with both the size and the overall $\sim B^2$ dependence of
the experimental data. 
The ratio of the experimental $C$ and calculated $C_N$  provides us
with an estimate for the fraction ($\xi$)
of Ga and As nuclei in the quantum wells that couple to the lattice. $\xi$ 
shows maxima of the order of unity at $\nu \approx 0.85$
and 1.2, and decreases as $\nu \rightarrow 1$ and for $\nu \gtrsim 1.2$
and $\nu \lesssim 0.85$. 
While the decrease in $\xi$ very near $\nu=1$ can be attributed to
the decreasing density of Skyrmions \cite{Brey95}, its decrease very 
far from $\nu=1$ (i.e. $\nu \gtrsim 1.2$ and $\nu \lesssim 0.85$) can 
be related to the 2DES approaching fillings where the Skyrmions are no 
longer relevant.
We note, moreover, that the $\nu$-dependences of $C$ and $\xi$ are
qualitatively similar to that of $\langle
S_{z} \rangle$ as deduced from the Knight-shift data \cite{Brey95,Barrett95};
in particular, both exhibit extrema at $\nu \approx 0.85$ and 1.2.

The temperature dependence of the high-$B$ heat capacity at $\theta
=0^{\circ}$ (Fig.~\ref{fig:1c}) and $\theta
=30^{\circ}$ (Fig.~\ref{fig:3}) at very low $T$ is particularly striking.
We observe that the $C \propto T^{-2}$ behavior is followed only down to 
$\sim$0.1K. For $T \lesssim$ 0.1K, $C$ increases faster with 
decreasing $T$ and, in a narrow range of $\nu$, $C$ exhibits a remarkably sharp 
peak at very low $T$ \cite{peakC}.
The deviation of $C$ from the $T^{-2}$ dependence at $T \gg \Delta 
/k_{B}$ and the shape of the observed peak are clearly not consistent with the 
Schottky model which predicts a
smooth maximum in $C$ at $T \sim \Delta /2 k_B$ ($\sim$2 mK in our case).
Instead, the shape and sharpness of this peak are
suggestive of a {\it phase transition} \cite{peak}.
The inset to Fig.~\ref{fig:3} shows that the peak temperature $T_c$ strongly 
depends on $\nu$ and $\theta$; in particular, $T_{c}$ 
decreases as $\nu \rightarrow 1$. 

It is tempting to associate the 
sharp peak in $C$ vs $T$ observed at low $T$ with the crystallization of 
Skyrmions, together with the associated magnetic ordering,
near $\nu$=1 which was recently proposed by Brey {\it et al.} 
\cite{Brey95}. The fact that our observed $T_c$ decreases as 
$\nu \rightarrow 1$ 
is consistent with the decreasing Skyrmion density which 
should reduce the Skyrme-solid melting temperature.
However, the details of the Skyrmion liquid-solid transition and, 
in particular, how it would affect the NSS are not known.
Here we remark on possible interpretations of our data.

First, we note that Eq.~\ref{CSchottky} alone cannot account for 
the observed anomaly: the NMR data \cite{Barrett95} suggest that 
$\Delta$ does not change significantly with decreasing $T$, and it is 
also expected \cite{Brey95,Girvin95} that a Skyrmion liquid-solid 
transition would affect $\Delta$ only very weakly.
We might therefore interpret the substantial enhancement of $C$ at low 
$T$ as an indication that 
either (1) more nuclei couple to the lattice, or (2) the
entropy of the coupled NSS decreases faster with decreasing $T$ than 
what is expected from the Schottky model. 
Picture (1) relies on a stronger coupling of the NSS to the lattice and 
also possibly an enhanced nuclear spin diffusion so that a larger number of 
nuclei contribute to the heat capacity near $T_c$. This is consistent 
with $T_c$ signaling the melting transition of Skyrmions: at such a 
phase transition, the coupling between the NSS and the electronic 
system is indeed expected to peak \cite{Girvin95}. 
Picture (2), on the other hand, relies on a Skyrme-solid induced nuclear spin 
polarization which reduces the entropy of the NSS.  
This is reminiscent of the dynamic nuclear polarization of the NSS,
for example when nuclear spins
interact with spin-polarized paramagnetic impurities \cite{Abragam78}.
A more relevant example is the induced polarization of the NSS in 
optically pumped NMR experiments, where the  nuclear spin polarization is much 
enhanced via their interaction with the spin-polarized 2DES 
\cite{Tycko95,Barrett95,Barrett94}.
While in the liquid Skyrme state motional narrowing prevents preferential 
orientation of electron spins, the transition to a pinned Skyrme solid 
could possibly induce a local preferential orientation of the 
electron-spin system which in turn would polarize the NSS and thereby 
reduce the entropy. We emphasize that these are possible 
interpretations; a definitive conclusion regarding the origin of the 
observed low-$T$ heat capacity anomaly awaits further experimental and 
theoretical work.

In conclusion, our heat capacity data underscore the importance of the 
coupling between the 2DES and nuclear spins, and point to the rich 
physics of the ground and excited states of the 2DES near $\nu$=1.

The authors are much indebted to S.M. Girvin, A.H. MacDonald and K.A. 
Moler for fruitful discussions and suggestions. M.S. thanks A. Kapitulnik for
warning him long ago (before the QHE Skyrmion days) about the possible
importance of the nuclear spins in GaAs/AlGaAs heat capacity measurements.
This work has been supported by NATO grant CRG 950328 and the NSF
MRSEC grant DMR-9400362.
V.B. acknowledges financial support of the Belgian National
Fund for Scientific Research.



\begin{figure}
\caption{Heat capacity $C$ at $\theta = 0^{\circ}$, showing orders of magnitude 
enhancement of the high-$B$ data (b) over the low-$B$ data (a).
The line through the data points is a guide to the eye.}
\label{fig:1}
\end{figure}

\begin{figure}
\caption{The temperature dependence of $C$ at $B$=7T ($\nu$=0.81) is 
shown in the main figure in a log-log plot. The dashed line 
shows the $T^{-2}$ dependence expected for the Schottky model. The 
inset shows a linear plot of $C$ vs $T$ at $B$=6.7T ($\nu$=0.85).}
\label{fig:1c}
\end{figure}

\begin{figure}
\caption{Heat capacity as a function of perpendicular magnetic
field $B_{\bot} = B \cos (\theta)$ and $\nu$ at $T$=100mK, and at the
indicated values of $\theta$. The curves correspond to the calculated
nuclear-spin
heat capacity of Ga and As atoms in the quantum wells $(C_{N})$ for $\theta =
0^{\circ}$ (dashed) and $30^{\circ}$ (solid).}
\label{fig:2}
\end{figure}

\begin{figure}
\caption{Temperature dependence of heat capacity at $\theta =
30^{\circ}$ and at the indicated values of $\nu$.
The $T^{-2}$ dependence expected for the
Schottky effect is shown as a dashed line, and $T_{c}$ is marked by
the vertical arrows. The $\nu$-dependence of
$T_{c}$ at $\theta =
0^{\circ}$ ($\circ$) and $30^{\circ}$ ($\bullet$) is shown in the inset
and the lines are guides to the eye.}
\label{fig:3}
\end{figure}


\begin{thebibliography}{10}
\bibitem{peakC} In the $T$-range ($T\lesssim 1.2 T_{c}$) where $C$ 
exhibits a maximum, we measured $C$ using the quasi-adiabatic 
technique, ensuring that the measured $C$ is not affected by possible changes 
in the internal time constant of the system.
\end{thebibliography}
\end{document}